\documentclass{emulateapj}
\usepackage[]{graphicx}

\newcommand{\chandra}{{\it Chandra }}

\begin{document}

\title{Evolution of X-ray spectra and light curves of V1494\,Aquilae}
\author{J.G. Rohrbach\altaffilmark{1}, J.-U. Ness\altaffilmark{1,2}, S. Starrfield\altaffilmark{1}}
\altaffiltext{1}{School of Earth and Space Exploration, Arizona
State University, Tempe, AZ 85287-1404, USA:
[Jonathan.Rohrbach, Jan-Uwe.Ness, Sumner.Starrfield]@asu.edu}
\altaffiltext{2}{European Space Astronomy Centre, PO Box 78, 28691 Villanueva de la Ca\~nada, Madrid, Spain: juness@sciops.esa.int}

\begin{abstract}
We present six Chandra X-ray spectra and light curves obtained for the nova V1494\,Aql (1999 $\#$2) in outburst. The first three observations were taken with ACIS-I on days 134, 187, and 248 after outburst. The count rates were 1.00, 0.69 and 0.53 cps, respectively. We found no significant periodicity in the ACIS light curves. The X-ray spectra show continuum emission and lines originating from N and O. We found acceptable spectral fits using isothermal APEC models with significantly increased elemental abundances of O and N for all observations. On day 248 after outburst a bright soft component appeared in addition to the fading emission lines. The Chandra observations on days 300, 304, and 727 were carried out with the HRC/LETGS. The spectra consist of continuum emission plus strong emission lines of O and N, implying a high abundance of these elements. On day 300, a flare occurred and periodic oscillations were detected in the light curves taken on days 300 and 304. This flare must have originated deep in the outflowing material since it was variable on short time scales. The spectra extracted immediately before and after the flare are remarkably similar, implying that the flare was an extremely isolated event. Our attempts to fit blackbody, Cloudy, or APEC models to the LETG spectra failed, owing to the difficulty in disentangling continuum and emission line components. The spectrum extracted during the flare shows a significant increase in the strengths of many of the lines and the appearance of several previously undetected lines. In addition, some of the lines seen before and after the flare are not present during the flare. On day 727 only the count rate from the zeroth order could be derived, and the source was too faint for the extraction of a light curve or spectrum.
\end{abstract}

\keywords{novae, cataclysmic variables –- stars: individual (V1494Aql) -- X-rays: stars -- methods: data analysis}

\section{Introduction}
When hydrogen-rich material is lost by a low-mass main sequence star and
accrets onto a white dwarf (WD) primary in a Cataclysmic Variable (CV), it settles onto the WD and eventually the bottom of the accreted layer becomes degenerate. When enough material has accumulated and the temperatures become high enough, a thermonuclear runaway is initiated and a Classical Nova (CN) outburst results. There is an initial short phase of X-ray emission which quickly fades as the ejecta expand and become optically thick. When the ejected shell expands and cools enough to become again transparent to X-rays, a soft, luminous X-ray source is typically observed, although, each CN evolves differently in X-rays. This phase of evolution in X-rays is called the SSS phase because X-ray spectra at this time resemble those of the class of super-soft X-ray sources (SSS; \citealt{kahab}).

V1494\,Aql was discovered in the optical by \cite{pereira99} on 1.785 December, 1999 at $m_{v}\cong6$ \citep{disc} and reached maximum light in the optical two days later, on 3.4 December, 1999 at $m_{v}\cong4.0$. It subsequently declined by two magnitudes in $6.6\,\pm\,0.5$ days thus classifying V1494\,Aql as a fast nova \citep{kissth00}. The distance to the nova was determined to be $1.6\pm0.2$ kpc by \cite{ii:1} and an orbital period of 0.13467 days has been suggested by \cite{orbit}. The hydrogen column density was estimated to $N_{\rm H}\approx4\times10^{21}$\,cm$^{-2}$ by \cite{ii:1} from sodium lines in the optical.

X-ray spectra were taken with {\it Chandra}, and \cite{krautter} reported that
the early evolution showed only emission lines, but that by Aug. 6, 2000 the spectrum had evolved into an SSS spectrum. They also reported an X-ray burst occurred (flare) and the presence of oscillations in one of the grating observations taken during the SSS phase. A detailed timing analysis was presented by \cite{drake:lc}.

We re-extracted all \chandra observations, and here we present the light curves and X-ray spectra. We carried out timing analyses, searching for periodic behavior in all observations. We present spectral modeling of the early observations that contain emission lines \citep{krautter} and provide a qualitative description of the SSS spectra. We also investigated spectral changes from spectra extracted before and after the flare event.


In the next section we present the observations and explain the extraction
techniques. We then focus on the timing analysis in \S\ref{timing} and the
spectral analysis in \S\S\ref{acis} and \ref{letg}. We discuss spectral
models in \S\ref{models} and summarize our results in \S\ref{disc}.

\section{Observations and Image Reduction}
\label{reduction}

\begin{table*}[!thb]
\begin{center}
\renewcommand{\arraystretch}{1.2}
\caption{\label{tab1}Observation log}
\begin{tabular}{lccccccc}
Start Date &Days After & Detector/& ObsId$^a$ & Exposure&Count Rate$^b$ & `Soft' Count Rate$^c$\\
    (UT)   & Outburst  &\ \ Grating &         & (ksec)  & photons/sec   & (photons/sec)     \\
\hline
2000, April 15, 01:01:27    &134&ACIS-I/none& 959     & 5.6  &1.0  & 0.12\\
2000, June 07, 02:48:14     &187&ACIS-I/none& 89      & 5.2  &0.69 & 0.07\\
2000, August 6, 22:02:05    &248&ACIS-I/none& 1709    & 5.6  &0.53 & 0.32\\
2000, September 28, 06:50:09&300&HRC-S/LETG & 2308    & 8.1  &0.65 & - \\
2000, October 1, 10:07:52   &304&HRC-S/LETG & 72      & 18.2 &0.84 & - \\
                            &   &           &pre-flare& 8.4  &0.71 & - \\
                            &   &           &flare    & 1.8  &3.17 & - \\
                            &   &           &post-flare&8.0  &0.74 & - \\
2001, November 28, 10:37:38 &727&HRC-S/LETG & 2681    & 25.8 &0.003& - \\
\hline
\multicolumn{7}{l}{$^a$ Observation Identification Number, $^b$ Zeroth Order count rate over entire bandpass, $^c$ Count rate for photons of energy}\\
\multicolumn{7}{l}{ less than 600 eV}
\end{tabular}
\renewcommand{\arraystretch}{1}
\end{center}
\end{table*}

We present six observations of V1494\,Aql taken with \chandra in 2000 and 2001. Table~\ref{tab1} gives the start date of the observations, the number of days since outburst, instrumental setup, observation identification number (ObsId), exposure time, net count rate and count rate for `soft' photons with an energy less than 0.6 keV. The first three observations were taken with the S-array of the Advanced CCD Imaging Spectrometer (ACIS-S), which is an array of CCD chips providing moderate spectral resolution in the energy range 0.2-10\,keV\footnote{http://cxc.harvard.edu/cdo/about\_chandra/\#ACIS}. After the SSS was detected with the ACIS observation taken on day 248 \citep{krautter}, the next observation used the High Resolution Camera (HRC-S) in combination with Low Energy Transmission Grating Spectrometer (LETGS), yielding higher spectral resolution. While the HRC detector has no energy resolution, the LETGS disperses the incoming light and projects a dispersed spectrum onto the HRC. LETG spectra are extracted in wavelength units (range 1-170\,\AA), but for consistency with the ACIS spectra we converted the LETG spectra to energy units.

We carried out the reduction with the \chandra-specific CIAO software suite, version 3.3. Since the CIAO standard data processing procedures (a.k.a. the 'pipeline') have changed since the time of the observations, we began our treatment of the images with the level 1 event files which were taken from the \chandra archives along with the accompanying calibration files. With the newest or most applicable calibration routines the exposures were reprocessed mimicking the pipeline reduction using standard routines from CIAO version 3.3.0.1 \citep{ciao}. These newly constructed event 2 files were used to create our light curves and spectra.

The light curves were extracted using tools developed by \cite{ness_lc} which determine point spread function (PSF) corrected source count rates. The PSF for each observation was constructed by following the CIAO threads and are specific to observation, detector location, and energy \citep{ciao}. The light curves extracted from the ACIS and the non-dispersed photons in the LETG observations (zeroth order) were extracted in 20 second time bins from a source extraction region of 20 pixels. This region encloses $99\%$ of the ACIS point spread function (PSF) and $98\%$ of the HRC PSF. The spectra were extracted from the level 2 event files following the CIAO threads. The ACIS spectra were extracted using a 20 pixel radius circular source extraction region and the LETG spectra were obtained from the combined $\pm1$ spectral orders.

In order to analyze the burst reported by \cite{drake:lc}, we applied a time filter to the data set taken on day 304 using the same time limits for the start and end time of the flare. This separation was done with the level 1.5 event files instead of the level 2 files so that the correct good time intervals could be applied. After this separation we had a pre-flare observation with an exposure time of 8.4 ksec, a 1.8 ksec exposure of the flare, and a post-flare exposure of 8.0 ksec. Each of these three event files were processed by the same reconstructed pipeline as described above to produce the new level two event files and spectra.

We expect pileup to not be a problem since the ACIS-S observations were designed to minimize this effect by placing the target approximately 7' from the aim-point and reducing the duration of each exposure to 0.8 sec. While PIMMS simulations show a small amount of pileup ($\sim20 \%$) this is an over-estimate since PIMMS does not model off-axis pointing.

\section{Photometry}
\label{timing}

\begin{figure}[!thb]
\begin{center}
\resizebox{\hsize}{!}{\includegraphics{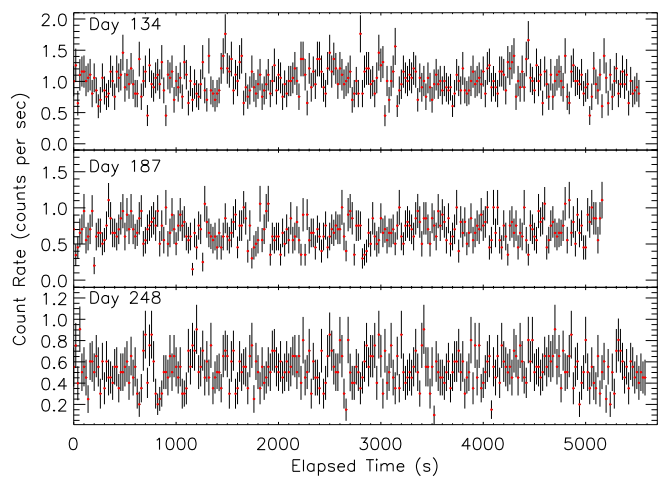}}
\caption{\label{lc1}Light curves for the ACIS observations taken on days 134, 187 and 248 plotted with error bars.}
\end{center}
\end{figure}

\begin{figure}[!thb]
\begin{center}
\resizebox{\hsize}{!}{\includegraphics{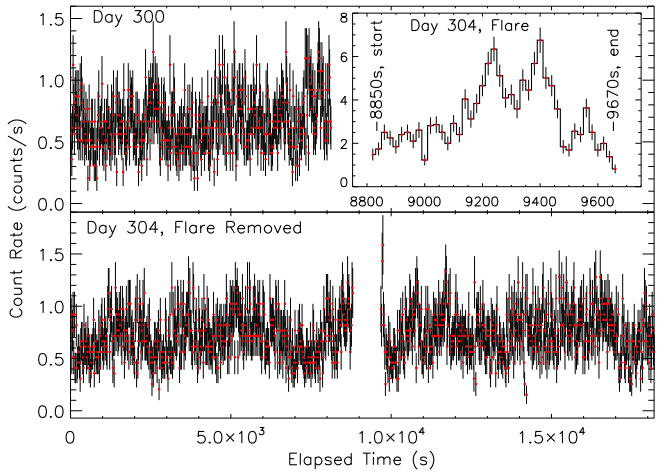}}
\caption{\label{lc2}Zeroth order light curves for LETG observations taken on days 300 and 304 plotted with error bars. The bottom panel shows the day 304 data with the flare removed while the inlay shows the light curve of the flare only. Note the difference in the vertical scale on the inlay.}
\end{center}
\end{figure}

\begin{figure}[!thb]
\resizebox{\hsize}{!}{\includegraphics{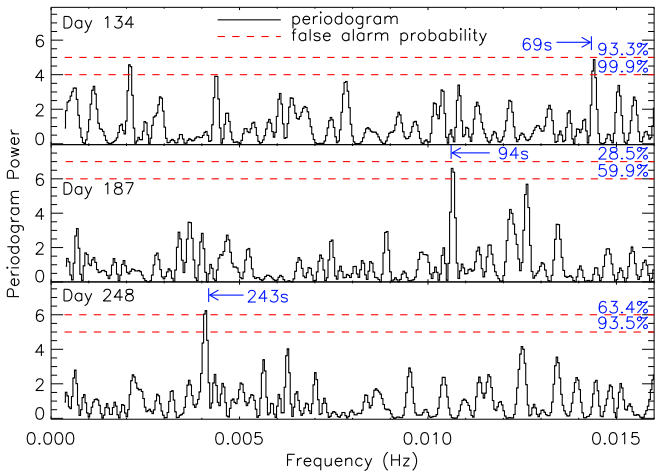}}
\caption{\label{acis_period}Periodograms for ACIS data. The horizontal dashed lines mark false alarm probabilities, and the peaks of the strongest signals are labeled, but none is statistically significant.}
\end{figure}

\begin{figure}[!thb]
\resizebox{\hsize}{!}{\includegraphics{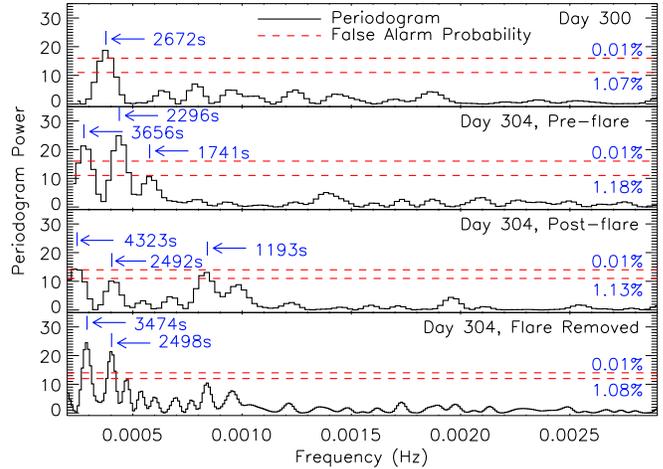}}
\caption{\label{letg_period}Periodograms for the HRC light curves. The dashed lines mark false alarm probabilities and the peak of the most significant signals are labeled. Note the difference in vertical and horizontal scales compared to the ACIS data.}
\end{figure}

The light curves for the ACIS observations are given in Fig.~\ref{lc1}, plotted with error bars. The ACIS data show the total count rate dropping as the nova evolves (see Table~\ref{tab1}). The count rate drops by 31\% between the first two observations. On day 248 the SSS spectrum emerged (\citealt{krautter}; see also Fig.~\ref{acis_energy}), but the total count rate had declined by 16\% compared to day 187.


While V1494\,Aql was detected on day 727 \citep{ness_lc}, only the count rate from the zeroth order could be determined, and the source was too faint for the extraction of a spectrum or reliable light curve. Thus, ObsId 2681 is not further considered.

The timing analysis was done using the methods of \cite{period} which constructs a periodogram, normalized by the total variance, through Fourier analysis which is suitable for evenly or unevenly spaced data. Normalizing by the total variance of the data allows one to calculate a `False Alarm Probability' \citep{scargle} to aid in identification of false period detections. We examined periods ranging from twice the bin size to half the observation length. Periodograms were constructed for each observation while several versions were analyzed for the observation containing the flare. We checked the data set from day 304, ObsId 72, for periodicity over the whole observation with the flare removed, only the pre-flare segment, and only the post-flare segment.

As seen in Fig.~\ref{acis_period}, none of the ACIS light curves show strong evidence for periodicity. The strongest signal corresponds to a false alarm probability of 42\% (day 187 at $\sim95$s). The LETGS data, on the other hand, show periods with false alarm probabilities well below 0.1\%, indicating a real signal, for each of the observations. In Fig.~\ref{letg_period} we show the periodograms of the HRC data with all peaks labeled whose signal strength relates to a false alarm probability less than 1\%. The flare is double peaked with additional structure. For further information the reader is directed to \cite{drake:lc}.

We detected signals close to the 2500s period reported by \cite{drake:lc} in all of the grating data. None of those data showed any significant evidence for periods shorter than $\sim1200$s (longer than $8.33\times10^{-4}$Hz), so this region is not plotted. For the data from day 300, the $\sim2500$s signal was the only one detected although the data from day 304 also showed several other strong features. The pre-flare periodogram shows features at $\sim1740$s and $\sim3600$s while the post-flare has signals at $\sim$1200s and $\sim4300$s. The 1200s signal could be spectral leakage from both the 3600s and 2500s signals while the 1740s feature could be leakage from the 3600s feature. We also searched for longer period signals in the entire exposure taken on day 304 with the flared portion removed. No longer period signals were seen but the 2500s and 3600s features were reproduced while neither the 1200s or 1740s features were present.

 We are unable to test for the presence of the orbital period of 3.23 hours
(=11.6\,ks) suggested by \cite{orbit}, because with our exposure times we
cannot sample any periods longer than 9\,ks.

The short duration of the flare, along with its photometric variability imply that
the source region of this flare must be associated with the hot WD. Variability on time scales of 200s (the width of one peak during the flare and 10 times our temporal bin size), yields a flare source region physical size of approximately 0.4 AU.
While this size is far larger than either a white dwarf or that of the binary
system, it is much smaller than the radius of the ejecta after 300 days of
expansion.

\section{ACIS spectra}
\label{acis}

\begin{figure}[!thb]
\begin{center}
\resizebox{\hsize}{!}{\includegraphics{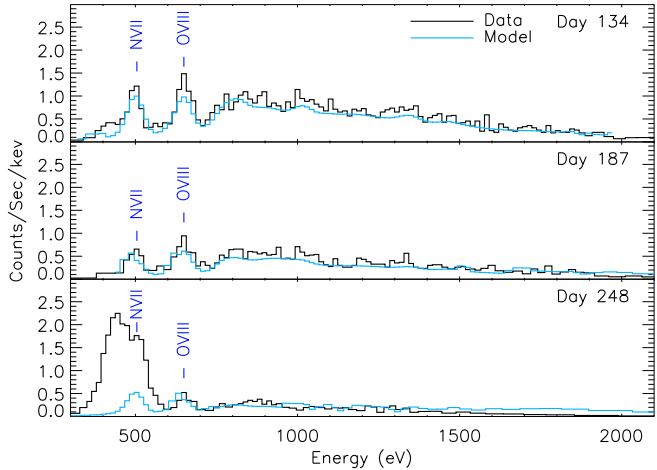}}
\caption{\label{acis_energy}ACIS-S spectra taken on the days indicated in the right legends. The emergence of the bright feature below 600 eV on day 248 (bottom panel) marks the beginning of the SuperSoft Source phase of this nova. The lines of O\,{\sc viii} and N\,{\sc vii} are labeled at their rest energies. We show APEC models fitted to the data with a light color (see discussion in section \ref{models}).}
\end{center}
\end{figure}

The first spectrum, taken on day 134 (top panel of Fig.~\ref{acis_energy}), shows the presence of N\,{\sc vii} and O\,{\sc viii} emission lines at 500 and 650 eV. The O\,{\sc viii} line is stronger than the N\,{\sc vii} line. In addition, continuum emission, unresolved lines, or the combination of both can be identified up to about 2 keV. By day 187, the lines of N\,{\sc vii} and O\,{\sc viii} as well as the continuum emission have weakened, but the relative strengths of the N\,{\sc vii} and O\,{\sc viii} lines seem unchanged. We see no evidence for the appearance of any new features and no excess emission at energies below 600 eV. The decline in the emission line strengths agrees with the reduction in count rate from the photometric data.

The spectrum from day 248 shows further weakening of the O\,{\sc viii} line and the possible emergence of a blend of Ne\,{\sc ix} lines at $\sim 900$\,eV. In addition, a new bright feature appears at energies just below the N\,{\sc vii} emission line at 500 eV. As seen in the photometry described in section \ref{timing}, the reduction in count rate is caused by the reduction of high-energy continuum emission. This reduction occurs in spite of the increase in count rate from below $\sim 600$ eV, probably caused by the appearance of the SSS \citep{krautter}. We tentatively interpret this feature as evidence that the density of the ejecta has declined, becoming optically thin to low-energy X-rays. We note, however, that the spectral resolution of the ACIS detector at low energies is limited by the quantum efficiency, and for studies of the soft component the LETGS is better suited.

\begin{table}[!thb]
\begin{center}
\renewcommand{\arraystretch}{1.5}
\caption{\label{tab2}Line Identification}
\begin{tabular}{cccl}
$\lambda$ & Energy  &    Possible    & Transition\\
  (\AA)   &   (eV)  & Identification & lower level - upper level\\
\hline
33.74	&	367.5	&	C\,{\sc vi}	&	1s $^2$S$_{1/2}$ - 2p $^2$P$_{1/2,3/2}$\\
29.53	&	419.9	&	N\,{\sc vi}	&	1s$^2$ $^1$S$_0$ - 1s2s $^3$S$_1$\\
29.08	&	426.4	&	N\,{\sc vi}	&	1s$^2$ $^1$S$_0$ - 1s2p $^3$P$_{1,2}$\\
28.79	&	430.7	&	N\,{\sc vi}	&	1s$^2$ $^1$S$_0$ - 1s2p $^1$P$_1$\\
28.47	&	435.6	&	C\,{\sc vi}	&	1s $^2$S$_{1/2}$ - 3p $^2$P$_{1/2,3/2}$\\
26.99	&	459.4	&	C\,{\sc vi}	&	1s $^2$S$_{1/2}$ - 4p $^2$P$_{1/2,3/2}$\\
26.36	&	470.5	&	C\,{\sc vi}	&	1s $^2$S$_{1/2}$ - 5p $^2$P$_{1/2,3/2}$\\
24.96   &   496.8   &   N\,{\sc vi} &   1s$^2$ $^1$S$_0$ - 1s3p $^3$P$_1$\\
24.90	&	498.0	&	N\,{\sc vi}	&	1s$^2$ $^1$S$_0$ - 1s3p $^1$P$_1$\\
24.78   &   500.4   &  N\,{\sc vii} &   1s $^2$S$_{1/2}$ - 2p $^2$P$_{1/2,3/2}$\\
23.79   &   521.2   &   N\,{\sc vi} &   1s$^2$ $^1$S$_0$ - 1s4p $^3$P$_1$\\
23.77	&	521.7	&	N\,{\sc vi}	&	1s$^2$ $^1$S$_0$ - 1s4p $^1P_1$\\
23.29   &   532.4   &   N\,{\sc vi} &   1s$^2$ $^1$S$_0$ - 1s5p $^3$P$_1$\\
23.28	&	532.7	&	N\,{\sc vi}	&	1s2 $^1$S$_0$ - 1s5p $^1P_1$\\
22.10   &   561.1   &  O\,{\sc vii} &   1s$^2$ $^1$S$_0$ - 1s2s $^3$S$_1$\\
21.80   &   568.5   &  O\,{\sc vii} &   1s$^2$ $^1$S$_0$ - 1s2p $^3$P$_{1,2}$\\
21.60	&	574.0	&  O\,{\sc vii} &	1s$^2$ $^1$S$_0$ - 1s2p $^1$P$_1$\\
18.97	&	653.6	& O\,{\sc viii}	&	1s $^2$S$_{1/2}$ - 2p $^2$P$_{1/2,3/2}$\\
16.01	&	774.7	& O\,{\sc viii}	&	1s $^2$S$_{1/2}$ - 3p $^2$P$_{1/2,3/2}$\\
\hline
\end{tabular}
\renewcommand{\arraystretch}{1}
\end{center}
\end{table}

\begin{figure}[!thb]
\begin{center}
\resizebox{\hsize}{!}{\includegraphics{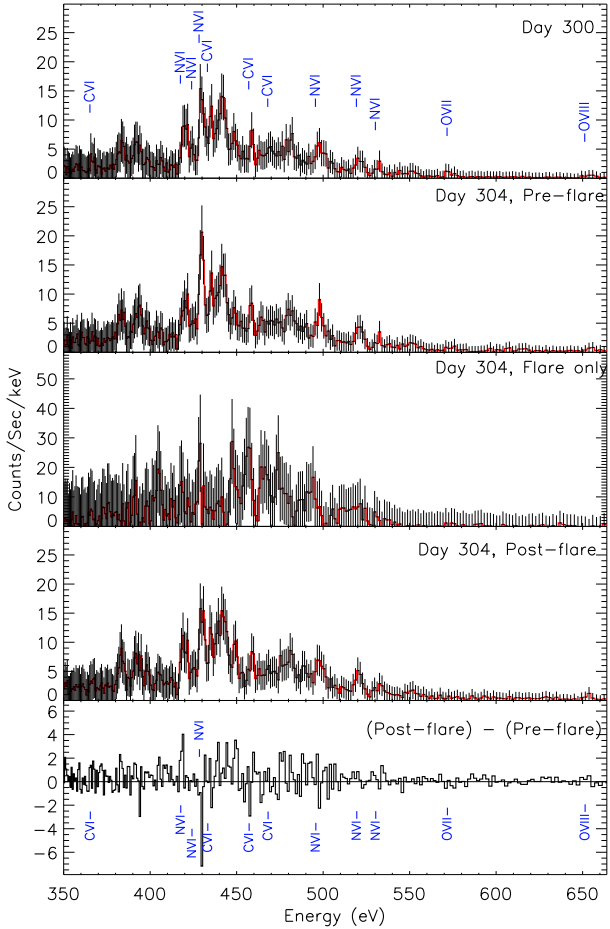}}
\caption{\label{letg_energy}\chandra HRC/LETG spectra with the most probable line identifications given. The bottom panel shows the difference between the spectra taken on day 304 before and after the flare event and is plotted without error bars for clarity.}
\end{center}
\end{figure}

\section{LETG spectra}
\label{letg}

Day 300 shows emission lines sitting on top of a continuum, see the top panel of Fig.~\ref{letg_energy}. This arrangement makes disentangling the continuum from the lines and line identifications difficult. As a result, we can only make qualitative arguments in this analysis. Table~\ref{tab2} gives a list of all the possible line identifications for the LETG data along with the associated wavelength, energy and atomic transitions (lower - upper states). Several unidentified lines exist between 400 and 500 eV, a characteristic that was also seen in the recent nova RS Oph as discussed by \cite{rsoph,rsophshock}.

We see lines from O\,{\sc viii}, C\,{\sc vi} and N\,{\sc vi} as well as several unidentified lines. While the line at $\sim 500$ eV could be N\,{\sc vii}, we interpret this line as N\,{\sc vi} due to the presence of the N\,{\sc vi} He-like $\gamma$ (1s-4p) and $\delta$ (1s-5p) lines and the absence of the N\,{\sc vii} $\beta$ (1s-3p) line at 640.5 eV. The strongest line in the spectrum taken on day 300, at 442 eV, is an unidentified line. There are also unidentified lines at $\sim$ 383, 393, 397, 404, 449 and 482 eV, as well as several other possible weak lines throughout the spectrum. The energy of the O\,{\sc viii} line (656 eV) is well above the Wien tail of the continuum ($\sim$ 530\,eV) and can therefore not be photoexcited. At least this one line must therefore be purely collisional. Since the O\,{\sc viii} line is much weaker than the lines at lower energies, it is reasonable to assume that the stronger lines are mixture of radiative and collisional excitations.

We constructed three sets of data from the observation made on day 304, as discussed in section \ref{reduction}, and their spectra are also presented in Fig.~\ref{letg_energy}. The second panel shows the pre-flare spectrum, the third shows the flare spectrum, and the fourth shows the post-flare spectrum, each plotted with error bars and labels for the most likely line identifications. The spectrum during the pre-flare period again shows emission lines on top of a continuum. However, there is stronger emission from N\,{\sc vi} at 430 eV and 498 eV but lines from O\,{\sc vii} at 571 eV and C\,{\sc vi} 367 eV were weaker than four days prior. The other lines, including all the unidentified lines, do not appear to significantly change in strength when compared to day 300.

In order to isolate the spectrum of the flare, we subtracted the count rate spectra from before and after the flare from the data recorded during the flare event. The bottom panel of Fig.~\ref{letg_energy} shows the difference spectrum from before and after the flare (in units of count rates per keV). This difference spectrum allows us to identify in which way the flare may have altered the emitting plasma in ways such as heating or photoionization. Examining the plot, we see that the flare event had little effect on the spectrum, yielding essentially the same spectrum as before the flare. This means that the difference between the pre-flare and post-flare count rate spectra yields the emission that originates from the plasma that emits the flare. This flare-only spectrum is shown in the third panel and an emission line spectrum with little or no sign of a continuum can be seen. During the flare the count rate in the 449 eV unidentified line, which was the strongest feature in the day 300 spectra, doubles and there are also increased count rates for the unidentified lines at 393 eV, N\,{\sc vi} at 430 eV, C\,{\sc vi} at 459 eV, and N\,{\sc vi} at 498 eV.  Also, unidentified lines appear at 405 and 475 eV that were not seen in any of the other spectra. The unidentified line at 442 eV is weak during the flare while the C\,{\sc vi} 435 eV line and the unidentified line at 482 eV are not seen in the flare at all.

After the flare most of the lines appear to return to near their initial pre-flare level. There is a slight increase in the strength of the N\,{\sc vi} 420 eV line as well as in the 383 and 393 eV unidentified lines but no strong changes are seen in any of the other N\,{\sc vi} lines. A significant portion of the total difference shown in the bottom panel of Fig.~\ref{letg_energy} comes in the wings of the unidentified lines at 442 and 449 eV. There is a small increase in emission on either side of these lines after the flare but there is no difference in the strength of the peaks. This may be due to some change in the continuum or an increase in the temperature. There is also a small amount of emission seen post-flare at 475 eV which must be residual from the flare. Unfortunately, the flare-only spectrum is not well-enough exposed to detect any changes in the continuum level, because the flare lasted only a short time \citep{drake:lc}.
If the flare-only spectrum is primarily an emission line spectrum, then
the flare could originate from the same regions that emit the emission
lines that blend with the continuum from the white dwarf. This would imply
that two physically distinct components are present. In view of the short
time scales of the flare, the regions that produce the emission line
component could be rather compact. 

 Another possibility could be that holes in the ejecta temporarily allowed
more ionizing continuum emission to reach the surrounding medium and
increase the degree of ionization. This would lead to stronger emission
lines, while some of the ionizing continuum emission could be radiating into a
different direction than the line of sight.
 We emphasize, however, that the data
are too limited for any strong conclusions, and these suggestions are
thus highly speculative.

\section{Spectral Models}
\label{models}

In this section we describe our attempts to find suitable spectral models.
We started with the ACIS spectra and fitted isothermal APEC models \citep{smith01}. The best-fit parameters for the models shown in Fig.~\ref{acis_energy} are given in Table~\ref{tab3} for each observation. In order to achieve satisfactory fits to the data, high abundances of N and O are required. While high abundances of N and O are not unusual in novae, the amount by which these elements has to be increased appears rather unrealistic (see, for example, the compilation of model predictions and observations of typical nova abundances listed by \citealt{JH98}, table 5). The best fit models require an oxygen abundance that is 20-30 times solar and a nitrogen abundance of several hundred times solar. We stress that, in addition to the quoted statistical uncertainties, there are significant sources of systematic uncertainties. We estimate the greatest source of uncertainty is the assumption of an isothermal plasma while the temperature structure of the emitting plasma is likely more complex, implying high- and low-temperature plasma. Since the O and N lines are formed at temperatures significantly below the temperatures of the isothermal model (see Table~\ref{tab3}), these lines are formed rather inefficiently. The only way to reproduce these lines and the high-energy emission at the same time is to increase the O and N abundances. A two-temperature model with a cool and a hot component yields lower N and O abundances (because the N and O lines are formed at low temperatures, while these elements are fully ionized at higher temperatures), however, such a model has more parameters and yields no significant improvement in reproducing the data. Another source of uncertainty is the parameter $N_{\rm H}$. Lower values of $N_{\rm H}$ require less emission at soft energies, thus leading to a higher resultant temperature and higher N and O abundances.

The model shown in the bottom panel of Fig.~\ref{acis_energy} (day 248) consists of
an APEC and a blackbody component, however, we only show the APEC component. We
found a good fit to the SSS component, yielding a combination of blackbody
temperature and luminosity (see footnote in Table~\ref{tab3}) that is typically
encountered when fitting blackbody curves to the SSS spectra of novae. The N
abundance is not constrained, owing to the overlap of the backbody component
with the N\,{\sc vii} line.

 Based on the blackbody parameters, we have tested an alternative model to the
observations taken on days 134 and 187. We have included a blackbody component
with the same parameters as those found for day 248. We kept the blackbody
parameters fixed and iterated only the APEC model parameters and $N_{\rm H}$.
We found the surprising result that better fits can be obtained, yielding
higher values of $N_{\rm H}$ ($4.4\times10^{21}$\,cm$^{-2}$ and
$5.5\times10^{21}$\,cm$^{-2}$, respectively) and a lower APEC temperature. The
abundances of N and O are also lower with the 2-component model. This result
allows the possibility that a SSS component could have been present all the
time and was only hidden behind a higher column of neutral hydrogen. However,
this exercise also demonstrates the sensitivity of models to such soft spectra
to the assumed amount of interstellar absorption. This leads to a large
systematic uncertainty that is difficult to assess.

\begin{table}[!tbh]
\begin{center}
\renewcommand{\arraystretch}{1.1}
\caption{\label{tab3}Model Best Fit Parameters}
\begin{tabular}{l|ccc}
Model       & \multicolumn{3}{c}{Days After Outburst}\\
Parameter    & 134 & 187 & 248$^a$ \\
\hline
k$T$ (eV)  & $630\,\pm\,40$ & $750\,\pm\,50$ & $607\,\pm\,34$\\
$\log(VEM)$ (cm$^{-3}$) & $55.7\,\pm\,0.1$ & $55.3\,\pm\,0.1$ & $55.23\,\pm\,0.02$ \\
$N_{\rm H}$ ($\times10^{21}$\,cm$^{-2}$)& $3.4\,\pm\,0.3$ & $2.4\,\pm\,0.3$ & $3.0\,\pm\,0.1$\\
O abundance ($\times $solar)& $29\,\pm\,6$ & $35\,\pm\,8$ & $17\,\pm\,3$ \\
N abundance ($\times $solar)& $840\,\pm\,150$ & $788\,\pm\,180$ & $<112$\\
$\chi^2_{\rm red}$    & 2.8 & 1.5 & 1.97 \\
\hline
\end{tabular}
$^a$Plus Blackbody with $T_{\rm bb}=31\,\pm\,2$ and $\log(L_{\rm bol})=37.7\,\pm\,0.4$
\renewcommand{\arraystretch}{1}
\end{center}
\end{table}


The LETG spectra are also extremely difficult to model. We first attempted blackbody fits, but we found no satisfactory fits. The problem is that no spectral range can be identified that is free from line emission. We then tested a number of Cloudy models with a wide array of parameters, but none gave satisfactory results. The combination of optically thick plus optically thin plasma emission makes the LETG spectra of V1494 Aql particularly challenging.
A promising approach is to use the PHOENIX atmosphere code, but models for spectra like the ones presented here do not yet exist. Optimization of the code to fit these data is in progress (vanRossum \& Hauschildt, priv. comm.).

\section{Discussion and Conclusions}
\label{disc}

 The evolution of V1494\,Aql underwent two distinct phases. The first phase was characterized by hard X-ray emission, dominated by emission lines.
Our APEC fits to the ACIS observations yielded satisfactory results, and the spectra imply that the ejected gas is an optically thin plasma in collisional
equilibrium. It is possible that the source of emission is shocks within the ejecta, as has been proposed by \cite{obrien94}.
Unfortunately, we cannot test the predicted decline in count rate, because the third ACIS observation was contaminated by the rising
SSS, leaving us only with the two observations on days 134 and 187. Linear interpolation between these observations suggests a decline rate of 2.1
counts per second per year. For a strong shock, the post-shock temperature $T_s$ is given by
\begin{equation}
T_s = \frac{3}{16} \frac{\bar m v_s^2}{\rm k}
\end{equation}
where k is Boltzmann's constant and $\bar m = 10^{-24}$\,g is the mean particle mass, including electrons \citep[see, e.g.,][]{bode06}. If the observed emission is induced by a shock, then the temperatures derived from the isothermal APEC models (Table~\ref{tab3}) correspond to a shock velocity of $700-800$\,km\,s$^{-1}$. In a multitemperature plasma, the hottest component determines the shock velocity (see, e.g., \citealt{bode06}). If our spectra allowed us to resolve more temperature components, the shock velocity derived from the hottest component would be slightly higher (see, e.g., \citealt{rsophshock}), but not by much, since the average temperature of the isothermal models is dominated by the hottest component. Since these observations were taken more than three months after outburst, significant deceleration will have taken place which explains why these velocities are lower than the early expansion velocities of $-1300$\,km\,s$^{-1}$ found by \cite{moro99}. However, with only two velocity values, we can not determine the power-law index for different scenarios.

The second phase is the SSS phase that started some time before day 248 after outburst and overlapped in time with the first phase. Fig.~\ref{letg_energy} shows that some residual emission from the first phase can still be recognized at high energies on days 300 and 304. This is a similar situation to that encountered for RS\,Oph, where the shock emission was still detectable at high energies while the SSS spectrum dominated at low energies \citep{ness_2}. While the ACIS-S spectrum taken on day 248 shows a typical SSS spectrum, the details revealed by the \chandra grating spectra are remarkably different from what SSS spectra are usually like. For comparison, the SSS Cal\,83 was observed with the same instrument, and \cite{lanz04} presented spectra that show continuum emission with absorption lines that can be fitted with atmosphere models, but little line emission can be seen. While V1494\,Aql was the first CN to have been observed with high spectral resolution in X-rays, later grating observations of novae during their SSS phase revealed spectra more similar to that of Cal\,83, e.g., V4743\,Sgr \citep{ness_4743} or RS\,Oph \citep{ness_2}. Those spectra can be fit with stellar atmospheres \citep{lanz04,petz05,rsoph}. In contrast, the second-most prominent SSS, Cal\,87, is also dominated by emission lines \citep[e.g.,][]{greiner04} and may be more similar to V1494\,Aql. Since Cal\,87 is an eclipsing binary, the viewing geometry may be an explanation for the different X-ray spectra. However, no spectral analysis of the grating spectra of Cal\,87 has been presented, likely owing to the same complications that we encountered.  

Our photometric analysis revealed that the first phase of the evolution shows no periodic oscillations, while the SSS phase is modulated by short-period oscillations. The absence of such oscillations in the early observations supports the notion that they originate from the WD, which supports the interpretation by \cite{drake:lc} that these are non-radial g$^+$ pulsations.

\acknowledgments

JGR and SS received partial support from NSF and NASA grants to ASU.
JUN gratefully acknowledges support provided by NASA through \chandra\ Postdoctoral
Fellowship grant PF5-60039 awarded by the \chandra\ X-ray Center, which is operated by
the Smithsonian Astrophysical Observatory for NASA under contract NAS8-03060.

\bibliographystyle{apj}
\bibliography{aql}

\end{document}